\documentclass[traditabstract,a4,10pt]{aa}

\usepackage{graphicx,amsmath}
\usepackage{natbib}

\newcommand{\ain}{a_{\rm in}}
\newcommand{\aout}{a_{\rm out}}

\newcommand{\elie}{{\mathbf E}}
\newcommand{\elih}{{\mathbf H}}

\newcommand{\asinh}{\,{\rm asinh \;} }
\newcommand{\atanh}{\,{\rm atanh \,} }
\newcommand{\elik}{{\mathbf K}}
\newcommand{\elipi}{{\mathbf \Pi}}        
\newcommand{\pdz}{\partial_Z}

\begin{document}

\title{A substitute for the singular Green kernel\\ in the Newtonian potential of celestial bodies}
\authorrunning{J.-M. Hur\'e et al.}
\titlerunning{A substitute for the singular Green kernel in the potential problems}
\authorrunning{J.-M. Hur\'e \and A. Dieckmann}

\author{Jean-Marc Hur\'e\inst{1,2} \and A. Dieckmann\inst{3}}


\institute{Universit\'e de Bordeaux, OASU, 351 cours de la Lib\'eration, F 33405 Talence
\and
CNRS, UMR 5804, LAB, 2 rue de l'Observatoire, BP 89, F 33271 Floirac\\
\email{jean-marc.hure@obs.u-bordeaux1.fr}
\and
ELSA, Physikalisches Institut der Universit\"at Bonn, Nussallee 12, D 53115 Bonn\\
\email{dieckman@physik.uni-bonn.de}}
\date{\today}
\date{Received 14/11/2011 / Accepted 01/03/2012}


\abstract{
The "point mass singularity" inherent in Newton's law for gravitation represents a major difficulty in accurately determining the potential and forces inside continuous bodies. Here we report a simple and efficient analytical method to bypass the singular Green kernel $1/|\vec{r}-\vec{r}'|$ inside the source without altering the nature of the interaction. We build an equivalent kernel made up of a ``cool kernel'', which is fully regular (and contains the long-range $-GM/r$ asymptotic behavior), and the gradient of a "hyperkernel'',  which is also regular. Compared to the initial kernel, these two components are easily integrated over the source volume using standard numerical techniques. The demonstration is presented for three-dimensional distributions in cylindrical coordinates, which are well-suited to describing rotating bodies (stars, discs, asteroids, etc.) as commonly found in the Universe. An example of implementation is given. The case of axial symmetry is treated in detail, and the accuracy is checked by considering an exact potential/surface density pair corresponding to a flat circular disc. This framework provides new tools to keep or even improve the physical realism of models and simulations of self-gravitating systems, and represents, for some of them, a conclusive alternative to softened gravity.}

\maketitle

\section{Introduction}

As a direct consequence of Newton's law for gravitation \citep{newton1760,kellogg29}, the potential of any continuous distribution of matter inside a volume ${\cal V}$ at a point  P$(\vec{r})$ of space is given by
\begin{equation}
\psi(\vec{r})=-G\iiint_{\cal V}{\frac{\rho(\vec{r}') d^3 \tau}{|\vec{r}-\vec{r}'|}},
\label{eq:psi}
\end{equation}
where $\rho(\vec{r}')$ is the mass density at P$'(\vec{r}') \in \cal V$, and $d^3 \tau$ is the elementary volume\footnote{This form holds in electrostatics, where $\rho$ is the density of electric charges, and the constant is $\frac{1}{4 \pi \epsilon_0}$ (instead of $-G$).}. In general, this is a three-dimensional (3D) converging integral. The presence of the Green kernel $1/|\vec{r}-\vec{r}'|$ is known to represent a difficulty in calculating $\psi$ everywhere inside and very close to ${\cal V}$ since this function diverges as $\vec{r} \rightarrow \vec{r}'$. This singularity is classically avoided by converting the Green function into an infinite series \citep[e.g.][]{kellogg29,ct99}. Although exact, series expansions suffer from a low convergence rate since these are alternating series; besides, the number of integrals increases linearly with the number of terms considered up to the truncation order. These problems collectively constitute a real practical difficulty \citep{clement74,stonenorman92}. The proper treatment of the singularity is the subject of a longstanding challenge. Shifting the P-grid and the P$'$-grid relative to each other or raising the numerical resolution around the singularity are the most natural techniques \citep{syc90}, but these are of limited efficiency. When possible, the separate treatment of the asymptotic form of the singularity gives very good results \citep[e.g.][]{ansorg03,hure05}, although this approach renders the global treatment somewhat complex. The difficulty can also be tackled by introducing a ``softening length'' \citep{he88,ars89}. This recipe | widespread in disc simulations | must however be seen as nothing but a crude approximation that cannot be used for accurate modeling under a certain scale. Whatever the prescription for the softening length, which is generally linked to the resolution or smallest physical length scale \citep[e.g.][]{hp09}, the inferred force field is globally weaker than in the Newtonian case, and the evolution and stability of gaseous systems is inevitably impacted in a non-trivial manner \citep{romeo98,sommer98,ars89,ez98}. The Poisson equation of course provides another other way to derive $\psi$ numerically \citep{kellogg29,durand64}. This approach requires accurate boundary or interior/matching conditions only accessible through Eq.(\ref{eq:psi}), and complex geometries are not always easy to manage \citep{gc01}.
 
In this paper, we present a new means to evaluate Eq.(\ref{eq:psi}) that avoids the singularity, and, at the same time, properly accounts for it. This is achieved by replacing the singular kernel by an equivalent and regular, two-term form, one term being the gradient of a new scalar potential\footnote{Some aspects of the theory of tensor potentials are developped by \cite{chandra73} in the context of rotating, self-gravitating fluids.}. This is the aim of Sections \ref{sec:splitting} and \ref{sec:hyperkernel}. Kernel equivalence is fundamental to preserving the Newtonian character of the interaction on all scales, and for any separation (in particular at long-range). This reformulation is designed to be efficient within sources, and is not expected to surpass usual methods outside sources. From a practical point of view, the potential easily becomes accessible as diverging kernels have disappeared from the volume integrals.

Although not specific to a given system of coordinates, the calculus is developed in cylindrical coordinates for which the equivalent kernel takes a nominal form, in particular under axial symmetry (this equivalent kernel is probably not unique; see Section \ref{sec:axi}). It can be applied as it is to all rotating gaseous/solid bodies (stars, discs, planets, asteroids, etc.) in either steady state or not, and for various applications \citep{dermott79,hachisu86III,baruteaumasset08}. A basic, six-step algorithm is reported in Section \ref{sec:numexp}, together with a numerical experiment using the most simple quadrature and differentiation rules. There is no special assumption about the distribution of matter in space (density field and geometry or shape), making the method general, and transposable to domains of physics other than gravitation. Some interesting perspectives are listed in the last section. 

\begin{figure}[h]
\includegraphics[width=9cm]{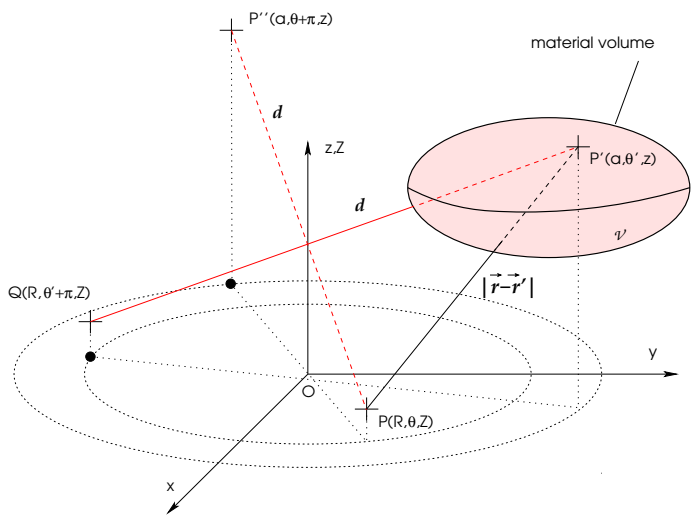}
\caption{Typical configuration for the gravitating, celestial body, and associated notations. See note \ref{notepq} for the definition of points P$''$ and Q.
}
\label{fig:scheme.eps}
\end{figure}

\section{Splitting of the Green kernel}
\label{sec:splitting}

We consider a volume of space ${\cal V}$ continuously filled with matter, as depicted in Fig. \ref{fig:scheme.eps}. Using cylindrical coordinates, with P'$(a,\theta',z)$ referring to source points, and P$(R,\theta,Z)$ to space points, the above integral for the Newtonian potential becomes
\begin{equation}
\psi(R,\theta,Z)= - G\iiint_{\cal V}{\frac{1}{\Delta}\rho(a,\theta',z) d^3 \tau},
\label{eq:psi_original}
\end{equation}
where $d^3 \tau = a da d\theta' dz$ is the elementary volume,
\begin{flalign}
\Delta^2 & = |\vec{r}-\vec{r}'|^2 \\
& = (a+R)^2+ \zeta^2 -4aR \sin^2 \phi,
\nonumber
\end{flalign}
\begin{equation}
\zeta=Z-z,
\end{equation}
is the relative altitude, and
\begin{equation}
2 \phi = \pi - (\theta - \theta').
\end{equation}

\begin{figure}
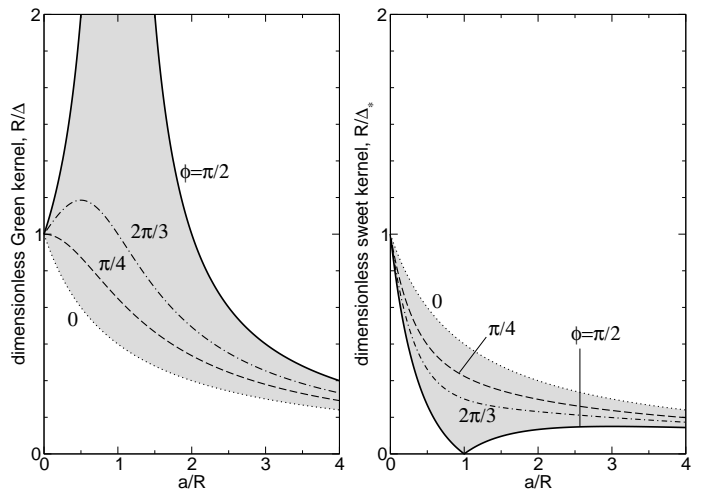

\includegraphics[width=4.5cm,bb=24 27 575 816,clip=]{aa18443_fig2.eps}\,\includegraphics[width=4.5cm,bb=24 27 575 816,clip=]{aa18443_fig3.eps}
\caption{Dimensionless Green kernel ({\it left}) and cool kernel ({\it right}) versus $a/R$ for $\zeta=0$ and $\phi=\{0,\frac{\pi}{4},\frac{2\pi}{3}, \frac{\pi}{2}\}$ labeled on the curves. The Green kernel diverges hyperbolically when $\vec{r} \rightarrow \vec{r}'$ (which corresponds to $a \rightarrow R$ and $\phi \rightarrow \frac{\pi}{2}$ here), in contrast to the cool kernel, which remains bounded (and even vanishes at the singularity).} 
\label{fig:kernels}
\end{figure}

Although $\Delta \rightarrow 0$ inside ${\cal V}$, the potential is generally a finite quantity \citep[i.e. integration is a regularizing process; e.g.][]{kellogg29,durand64}. We now set
\begin{equation}
\delta = \sqrt{(a+R)^2+\zeta^2} \ge R.
\end{equation}
This quantity is finite everywhere, and non-zero except on the polar axis (see below). Assuming $R>0$, we have
\begin{flalign}
\frac{\Delta}{\delta^2} & = \frac{1}{\Delta}\frac{\Delta^2}{\delta^2} \\ \nonumber
                         & =\frac{1}{\Delta} \left (1 - \frac{4aR\sin^2 \phi}{\delta^2} \right),
\end{flalign}
and so
\begin{equation}
\frac{1}{\Delta} = \frac{1}{\Delta_\star} + 4aR\sin^2 \phi \times \frac{1}{\Delta\delta^2},
\label{eq:ksplitting}
\end{equation}
where we have defined
\begin{equation}
\frac{1}{\Delta_\star} \equiv \frac{\Delta}{\delta^2}.
\label{eq:gentlekernel}
\end{equation}

The Green kernel is then split into two terms. The term $\frac{1}{\Delta_\star}$ is always regular\footnote{It is the inverse of a distance, and its value can be interpreted geometrically by noting that (see Fig. \ref{fig:scheme.eps})
\begin{equation}
\frac{\Delta}{\delta^2} = \frac{|\vec{r}-\vec{r}'|}{{\rm QP'}^2} = \frac{|\vec{r}-\vec{r}'|}{{\rm PP''}^2} =  \frac{|\vec{r}-\vec{r}'|}{d^2} < \infty,
\end{equation}
where Q$(R,\theta'+\pi,Z)$ is a point of space, diametrically opposite to P$'$, and P$''(a,\theta +\pi,z)$ is a point, diametrically opposite to P, that belongs to a fictitious source. \label{notepq}}
; we call this term the {\it cool kernel} in the following. Figure \ref{fig:kernels} displays $R/\Delta$ and $R/\Delta_\star$ versus $a/R$ around the singularity. We clearly see that the amplitude of $\Delta_\star$ is bounded, in contrast to $\Delta$. The second term in Eq.(\ref{eq:ksplitting}) is still singular when $\Delta \rightarrow 0$, but its integration over the material volume is expected to produce a regular field. The idea is to generate this singular kernel from the gradient of a regular function $\kappa$ (hereafter called {\it {hyperkernel}}), and to integrate it over the volume ${\cal V}$. As the two spaces $(R,\theta,Z)$ and $(a,\theta',z)$ are independent, the derivative may be drawn before the integral. This reasoning can be summarized as
\begin{flalign}
\text{singular kernel } & \equiv \nabla \text{ regular hyperkernel} \nonumber \\
&\downarrow \nonumber  \\
\iiint_{\cal V}{\text{sing. kernel } d^3 \tau}& = \nabla \iiint_{\cal V}{\text{reg. hyperkernel } d^3 \tau}, \nonumber
\end{flalign}
where the gradient  $\nabla$ is to be taken with respect to one of the three variables $R$, $\theta$, or $Z$. There are then three possible hyperkernels.

The existence of the hyperkernel is not guaranteed a priori, and it is of interest only if it is available in a closed form. The investigation indeed shows that the nominal form is obtained by considering the vertical gradient (i.e. $\nabla \equiv \pdz$). This may be due to the special role that the $Z$-axis plays a in cylindrical coordinates. We therefore do not discuss in detail any of the other two options, although these might be useful in certain circumstances.

\section{The singular term as the vertical gradient of a hyperkernel}
\label{sec:hyperkernel}

To get the hyperkernel, we consider the integration of the singular term in Eq.(\ref{eq:ksplitting}) with respect to $Z$. By using the intermediate variable $t=\zeta/\Delta$, we find after some algebra
\begin{flalign}
\int\frac{dZ}{\delta^2 \Delta} & = \frac{1}{(a+R)^2}\int{\frac{dt}{1-m^2 \sin^2\phi \times t^2}},
\end{flalign}
where
\begin{equation}
m=\frac{2\sqrt{aR}}{a+R},
\end{equation}
with $0 \le m \le 1$. We finally get
\begin{flalign}
\label{eq:zhyperkernel}
\kappa  & \equiv 4aR \sin^2 \phi  \int\frac{dZ}{\delta^2 \Delta} \\
& =  m \sin \phi \atanh \left( \frac{\zeta m \sin \phi }{\Delta} \right).
\nonumber
\end{flalign}
We could obviously add to $\kappa$ any function of $a$, $R$, $\theta$ and $\theta'$, but this is not necessary here as we take its $Z$-gradient. The Green kernel is then given by the equivalent form
\begin{equation}
\frac{1}{\Delta} = \frac{1}{\Delta_\star} + \pdz \kappa.
\end{equation}
It is necessary to verify that $\kappa$ is regular. This is straightforward since $\Delta > |\zeta|$ as soon as $R>0$. In other words, if $m \rightarrow 1$ and $\phi \rightarrow \frac{\pi}{2}$, then $\zeta/\Delta \rightarrow \pm 1$.

If we now multiply Eq.(\ref{eq:ksplitting}) by $\rho(a,\theta',z) d^3 \tau$ | which does not depend on $Z$ | and integrate over the material volume  ${\cal V}$, we find that
\begin{flalign}
\label{eq:newpsi-Z}
\iiint_{\cal V}{\frac{1}{\Delta}\rho d^3 \tau} & =  \iiint_{\cal V}{ \frac{1}{\Delta_\star}\rho  d^3 \tau} + \pdz \iiint_{\cal V}{ \rho \kappa  d^3 \tau},
\end{flalign}
where the partial derivative now operates on the integral. Up to a factor $-G$, this expression is precisely the potential defined by Eq.(\ref{eq:psi}), and it is both exact and general. It depends on neither the body's shape nor on the distribution of its mass density. It applies not only to volume distributions, but also to surface distributions (see Sect. \ref{sec:accuracy}) and linear distributions. The first integral in Eq.(\ref{eq:newpsi-Z}) is then the {\it cool potential} associated with the cool kernel, and the second term is the vertical gradient of a {\it hyperpotential}.

On the polar axis (i.e. $R=0$) $\Delta = \delta$, and so Eq.(\ref{eq:ksplitting}) does not help us to treat the singularity when $\zeta =0$ and $a =0$. In this case, we have
\begin{flalign}
\label{eq:hyperkernel_axis}
\kappa & \equiv \int{\frac{1}{\Delta}dZ}\\
& = \asinh \frac{\zeta}{a},
\nonumber
\end{flalign}
provided that $a>0$. The potential can then be written
\begin{equation}
\psi(0,Z)= -G \pdz \iiint_{\cal V}{\rho \kappa d^3\tau},
\label{eq:newpsi-Zontheaxis}
\end{equation}
where here there is no cool kernel. Besides, we see that $\lim_{a \rightarrow 0}~a \kappa~=~0$.

There is no continuity between the two different expressions for the cool kernel, the one valid at $R=0$ and the other valid at $R \rightarrow 0$ (this is also true for the hyperkernel). This is no problem as long as we do not have to consider the radial gradient of $\kappa$ (see below the numerical experiment).

Finally, we note that, as we work with an equivalent form of the Green kernel, the potential found from Eq.(\ref{eq:newpsi-Z}) automatically has  the right asymptotic property, and varies like $M/r$ sufficiently far away from the body. At large relative distance (i.e. $R \gg a$ and $Z \gg z$), we have $\delta \rightarrow r = \sqrt{R^2+Z^2}$. At the lowest order, one finds that 
\begin{flalign}
\iiint_{\cal V}{\frac{1}{\Delta_\star}\rho d^3 \tau}
 & \approx \frac{M}{r},
\end{flalign}
which means that the long-range behavior is exclusively contained in the cool kernel. 

\section{The case of axially-symmetric bodies}
\label{sec:axi}

Axially-symmetric bodies constitute an important class of astrophysical objects \citep{chandra73,hachisu86}. Interestingly enough, in problems where $\partial_{\theta'} \rho =0$, we can rewrite the above expressions in a more compact form in terms of elliptic integrals. The first integral in the right-hand-side of Eq.(\ref{eq:newpsi-Z}) becomes\footnote{A factor of two is due to $d\theta'/d\phi$, and another factor of two contained in the modulus $k$ comes from symmetry consideration (i.e. matter located at $\theta' \in [ \theta,\theta + \pi]$ provides the same contribution as matter located at $\theta' \in [ \theta,\theta - \pi]$).}
\begin{equation}
\label{eq:integwithe}
\iiint_{\cal V}{ \frac{1}{\Delta_\star}\rho  d^3 \tau}  = 2 \iint_{\cal S}{\rho \sqrt{\frac{a}{R}} k \elie(k)da dz},
\end{equation}
where
\begin{equation}
\elie(k)=\int_0^{\pi/2}\sqrt{1-k^2 \sin^2 x} \, dx,
\end{equation}
is the complete elliptic integral of the second kind, $k~=~2\sqrt{aR}/\delta$ is the modulus (with $k \in [0,1]$), and the double integration runs over the meridional cross section ${\cal S}$ of the body. The integration over the polar angle $\theta'$ of the hyperkernel gives:
\begin{flalign}
 \nonumber
 & \int_\phi{ \sin \phi \atanh \left( \frac{\zeta m \sin \phi }{\Delta} \right) d\phi}  = -  \cos \phi \atanh \left( \frac{\zeta m \sin \phi }{\Delta} \right) \\
 & \qquad \qquad +  \frac{\zeta }{m\delta}  \left[ F(\phi,k) - {m'}^2 \Pi(\phi,m,k) \right],
\end{flalign}
where
\begin{equation}
F(\phi,k) = \int_0^{\phi}\frac{d x}{\sqrt{1-k^2\sin^2 x}},
\end{equation}
is the incomplete elliptic integral of the first kind, and
\begin{equation}
\Pi(\phi,m,k) = \int_0^{\phi}\frac{d x}{\left(1-m^2\sin^2 x \right) \sqrt{1-k^2\sin^2 x}},
\end{equation}
is the incomplete elliptic integral of the third kind ($m$ is the parameter and $m' = \sqrt{1 - m^2}$). Over the whole circle (i.e. $\phi \in [0, \frac{\pi}{2}]$), this yields the axially symmetric potential\footnote{If we perform the $Z$-derivative and rearrange terms, we recover the well-known expression \citep{durand64}, namely
\begin{equation}
\psi(R,Z)=-2G\iint_{\cal S}{\rho \sqrt{\frac{a}{R}} k \elik(k) da dz },
\end{equation}
whose kernel is logarithmically singular.}:
\begin{flalign}
\label{eq:newpsi2_zkernel}
\psi(R,Z) = & -2G \iint_{\cal S}{\rho \sqrt{\frac{a}{R}} k \elie(k) da dz } \\ \nonumber
& \qquad -2G \pdz \left[ \iint_{\cal S}{\rho \sqrt{\frac{a}{R}} \zeta \elih(m,k) } da dz  \right],
\end{flalign}
where $\elih$ is defined for convenience by
\begin{equation}
\elih(m,k) = k \left[ \elik(k) - {m'}^2 \elipi(m,k) \right],
\end{equation}
with $\elipi (m,k)= \Pi(\frac{\pi}{2},m,k)$ and $\elik(k)=F(\frac{\pi}{2},k)$. The presence of the $\elih$-function is actually expected here since $\partial_z \Delta$ and $\partial_Z \Delta$ are linked \citep{trovahh12}.

On the polar axis, we get
\begin{equation}
\psi(0,Z)= - 2\pi G \pdz \iint_{\cal S}{\rho  \kappa a da dz},
\label{eq:newpsi2-Zontheaxis}
\end{equation}
where $\kappa$ is, in this case, given by Eq.(\ref{eq:hyperkernel_axis}). We note that, in this axially symmetric case, the potential could be determined through Eq.(\ref{eq:integwithe}) (i.e. by using the cool kernel only, and no hyperkernel), but the integrand still contains a hyperbolic divergence as $a \rightarrow 0$ and $\zeta \rightarrow 0$ which is not easy to manage. This is why it seems much better to consider Eq.(\ref{eq:newpsi2-Zontheaxis}), as the logarithmic divergence of the hyperkernel (i.e. the $\asinh$ term) is cancelled out by the elementary volume when $a$ is close to $0$ (i.e. $\lim_{a \rightarrow 0}~a \kappa~=~0$).

\begin{figure*}
\begin{center}
\includegraphics[width=7.5cm,bb=122 70 340 277,clip=]{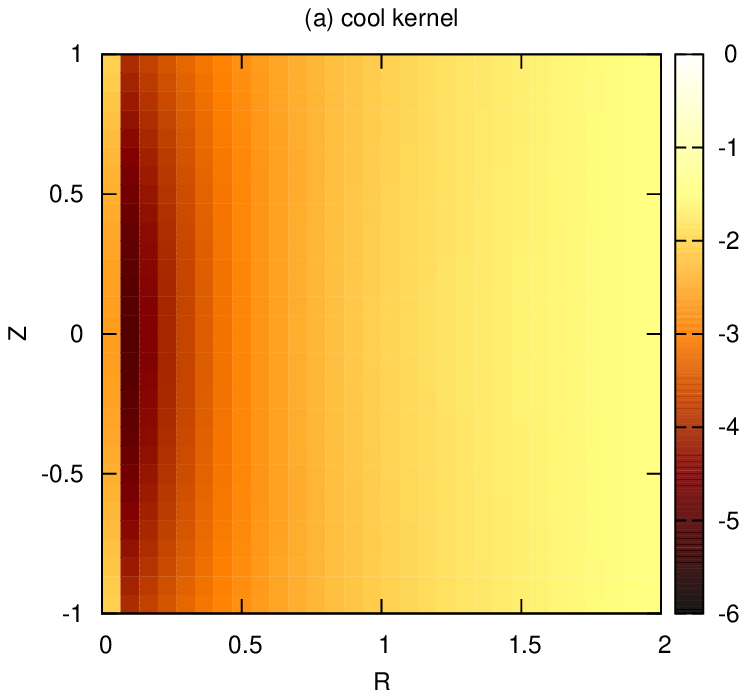}
\includegraphics[width=7.5cm,bb=122 70 340 277,clip=]{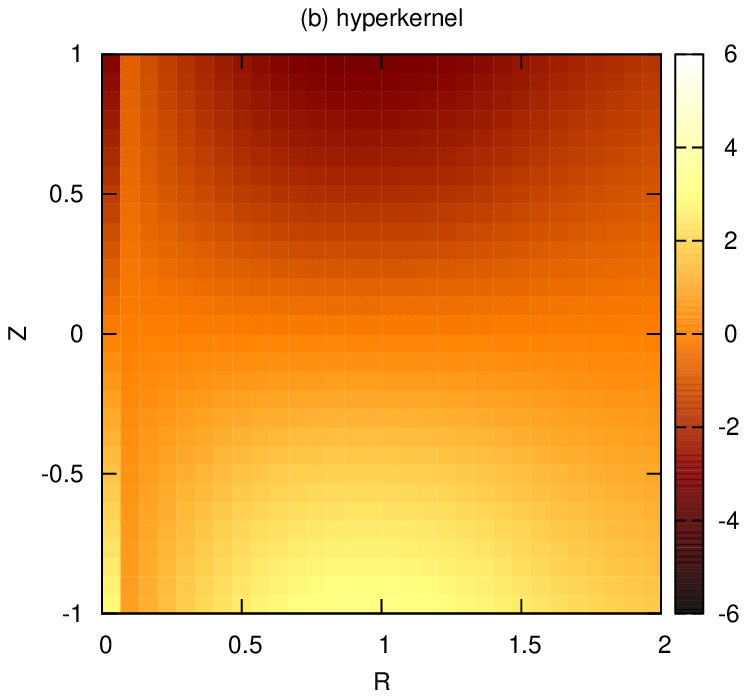}
\includegraphics[width=7.5cm,bb=122 70 340 277,clip=]{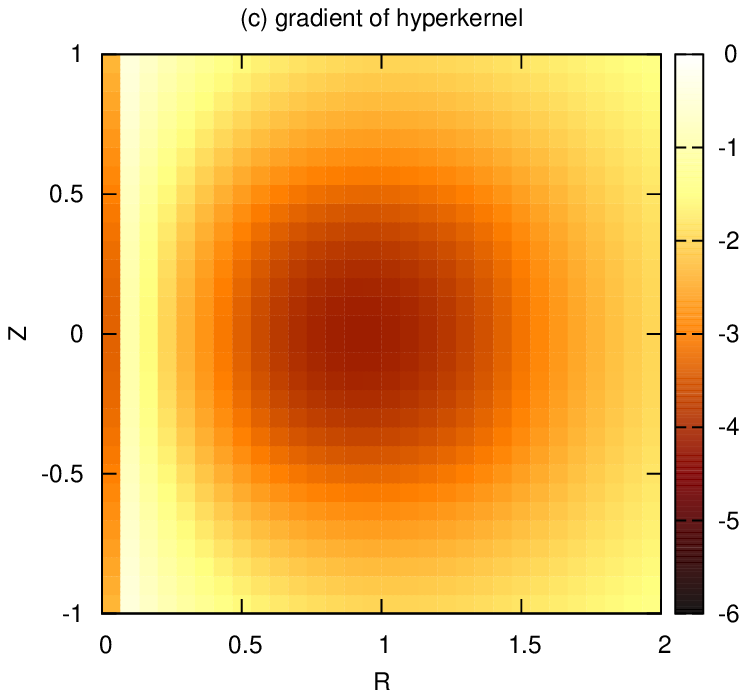}
\includegraphics[width=7.5cm,bb=122 70 340 277,clip=]{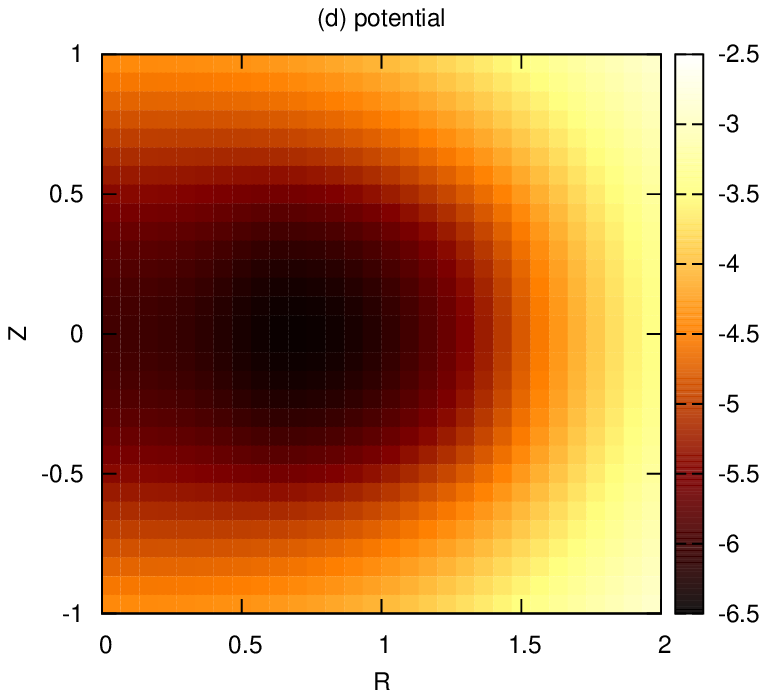}
\end{center}
\caption{Main steps in the computation of the potential from Eqs.(\ref{eq:newpsi2_zkernel}) and (\ref{eq:newpsi2-Zontheaxis}) in a typical case: (a) the integral of the cool kernel, (b) the integral of the hyperkernel,  (c) its vertical gradient, and (d) the potential as the sum of maps (a) and (c). Here, the body is an axially symmetric torus with square cross section (boundary indicated with a white line). As clearly visible in graphs (a)-(c), the treatment differs slightly on the polar axis (first column of pixels). See the text for the numerical setup.
}
\label{fig:res}
\end{figure*}

\section{An example of implementation}
\label{sec:numexp}

We now present a first numerical experiment to briefly describe the main steps of the method, and demonstrate its simple implementation. For a full $3$D body, the main steps\footnote{Step 1 should be executed once and for all, except for systems that evolve with time.}
 of any numerical estimate can be summarized as
\begin{enumerate}
\item Discretize the material source ${\cal V}$ on a 3D-grid, in the form of quadruplets $\{{\rm P'}(a_{i,j,k},\theta'_{i,j,k},z_{i,j,k}),\rho_{i,j,k}\}$,
\item Select a point P$(R,\theta,Z)$ in space,
\item For each point P'$_{i,j,k}$ of the source, compute the cool kernel, and the hyperkernel $\kappa$ from Eqs.(\ref{eq:gentlekernel}) and (\ref{eq:zhyperkernel}) or Eq.(\ref{eq:hyperkernel_axis}),
\item Perform the volume integrals in Eq.(\ref{eq:newpsi-Z}) or (\ref{eq:newpsi-Zontheaxis}),
\item Determine the vertical gradient of the hyperpotential,
\item Add this derivative to the cool potential, and multiply by $-G$,
\end{enumerate}
and reiterate steps $2$ to $6$ to generate a potential map. The components of the associated gravitational force are deduced as usual from the three gradients of $\psi$.

We have considered a homogeneous, axially symmetric torus with a square meridional cross section ${\cal S}$ with $(a,z) \in [\frac{1}{2},\frac{3}{2}] \times [-\frac{1}{2},\frac{1}{2}]$. The mass density ($\rho=1$ inside ${\cal S}$ and $0$ outside) is defined on a regular mesh, with $N$ nodes in each direction. The computational grid also consists of a square box $(R,Z) \in [0,2] \times [-1,1]$ with $L$ nodes per direction and regular spacing, therefore encompassing the body. The double integrals in Eqs.(\ref{eq:newpsi2_zkernel}) and (\ref{eq:newpsi2-Zontheaxis}) are computed at each node of the computational grid through the two-point trapezoidal rule. The partial derivative of the hyperpotential is estimated through second-order finite differences. These basic schemes are very easy to implement and the above six-step procedure contains no pitfall. We take $N=L=31$ here. Figures \ref{fig:res}a-d display, respectively, the integral of the cool kernel, the integral of the hyperkernel $\kappa$, its vertical gradient, and the total potential, obtained by adding the first and third maps. The boundary of the toroidal body is superimposed on these maps. We note that the vertical gradient of the hyperpotential makes the geometrical cross section rise above the background, and filters the curvature effects (which are enhanced by the integral of the cool kernel).

The computing time is typical of integral methods. Under the conditions of the present example, the integration of the cool kernel and hyperkernel (step 4 above) requires $ 2 N^2$ elementary operations per point of space. To get the potential at the nodes of a $L \times L$ square grid, we then find $2 N^2L^2$ (the time needed to determine the vertical gradients is negligible in comparison). This is therefore much smaller than that is usually obtained from an expanded Green function by a factor equal to the number of terms up to the truncation order \citep[which can be as high as a few hundred; see e.g.][]{hachisu86,stonenorman92}.

\begin{figure*}
\begin{center}
\includegraphics[width=7.5cm,bb=122 70 340 277,clip=]{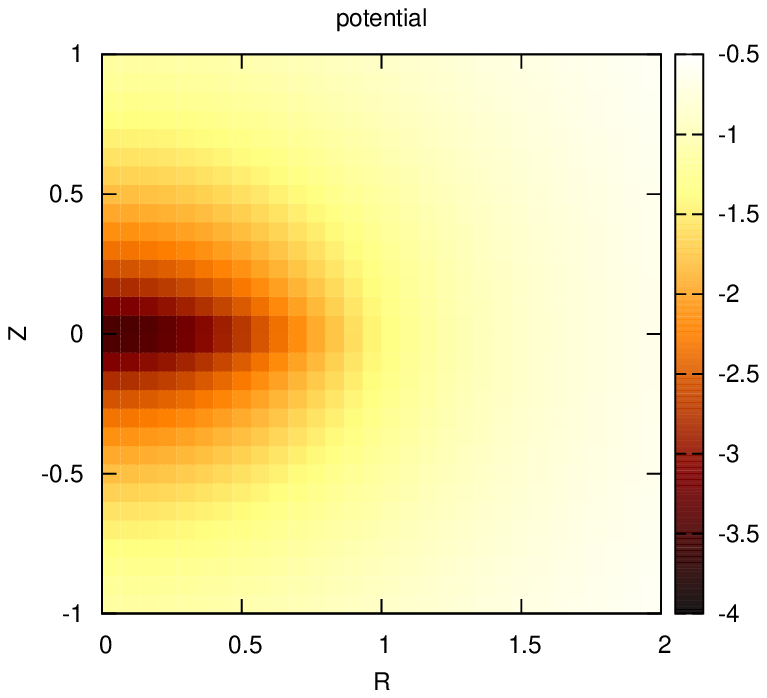}
\includegraphics[width=7.5cm,bb=122 70 340 277,clip=]{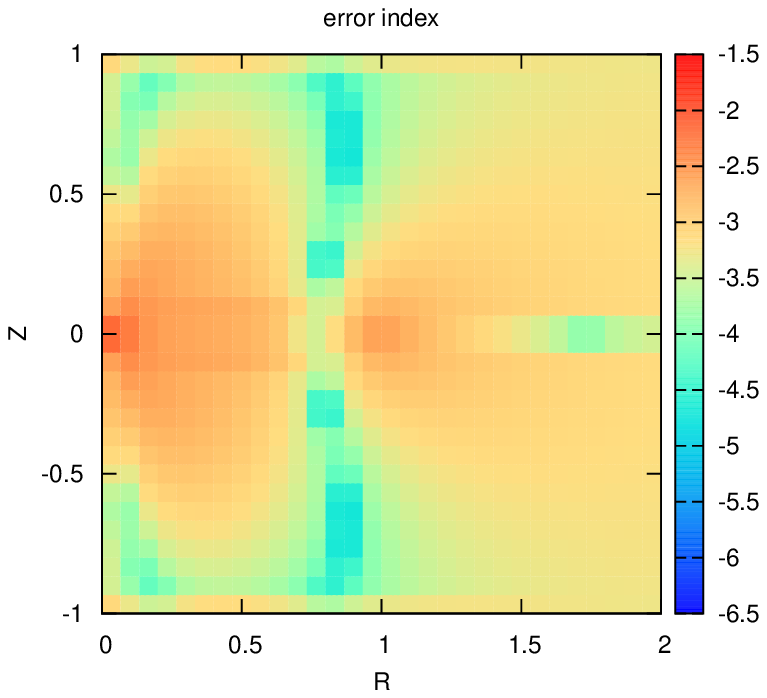}
\end{center}
\caption{Same legend as for Fig. \ref{fig:res} but for the flat, axially symmetric disc with surface density $\Sigma \propto (1-a^2)^{3/2}$, inner edge $\ain=0$ and outer edge $\aout=1$. The disc is indicated with a white line. The exact potential $\psi_e$ ({\it left}) is derived from the formula of \cite{schulz09}. The associated error index ({\it right}) is determined once $\psi$ is computed from Eqs.(\ref{eq:psi_flat_zkernel}) and (\ref{eq:psi_flat-Zontheaxis}). The mesh size is $\frac{1}{30}$ corresponding to $N=31$ radial points.
}
\label{fig:res_ml}
\end{figure*}

\section{Checking the accuracy}
\label{sec:accuracy}

The numerical accuracy is sensitive to various ingredients, such as the quadrature and differentiation schemes. To a lesser extent, it also depends on the mass density distribution $\rho$ and equation of the boundary $\partial {\cal V}$, which may generate additional difficulties in the calculations (interpolation of data points, infinite derivatives at edges, etc.). We present a second numerical test illustrating the accuracy of the method by considering an exact potential/density pair. Not many configurations correspond to finite mass and finite size systems \citep[e.g.][]{binneytremaine87}. When matter is gathered in a plane (i.e. a flat disc), Eq.(\ref{eq:newpsi-Z}) is directly transposable by setting $\rho(a,\theta',z)=\Sigma(a,\theta') \delta(z)$ and integrating over $z$. Under axial symmetry, we respectively get\footnote{These two formulae can be compared with the classical expression \citep{durand64,binneytremaine87}:
\begin{equation}
\psi(R,Z) =  -2G \int_{\ain}^{\aout}{\Sigma \sqrt{\frac{a}{R}} k_0 \elik(k_0) da },
\end{equation}
which is singular inside the source.} from Eqs.(\ref{eq:newpsi2_zkernel}) and (\ref{eq:newpsi2-Zontheaxis})
\begin{flalign}
\label{eq:psi_flat_zkernel}
\psi(R,Z) = & -2G \int_{\ain}^{\aout}{\Sigma \sqrt{\frac{a}{R}} k_0 \elie(k_0) da } \\ \nonumber
& \qquad -2G \pdz \left[ \int_{\ain}^{\aout}{\Sigma \sqrt{\frac{a}{R}} Z \elih(m,k_0) } da \right],
\end{flalign}
for $R>0$, and
\begin{equation}
\psi(0,Z)= - 2\pi G \pdz \int_{\ain}^{\aout}{\Sigma  \asinh \left(\frac{Z}{a}\right) a da},
\label{eq:psi_flat-Zontheaxis}
\end{equation}
where $\ain$ and $\aout$ denote the discs inner and outer edges, and
\begin{equation}
k_0^2=\frac{4aR}{(a+R)^2+Z^2}.
\end{equation}

By setting $\ain=0$ and $\Sigma \propto (1-a^2/\aout^2)^{3/2}$ for $a\in[0,\aout]$, one gets one of the three cases analyzed by \cite{schulz09}. For such a distribution, the associated potential, $\psi_e$, is known exactly in a closed-form for any point of space. Figure \ref{fig:res_ml} gives $\psi_e$ as well as the error index $\epsilon = \log_{10}|1-\psi/\psi_e|$ where $\psi$ is determined from Eqs.(\ref{eq:psi_flat_zkernel}) and (\ref{eq:psi_flat-Zontheaxis}) in the same conditions as above (we set $\aout=1$). We see that the potential outside and especially inside the disc is well-reproduced. The relative error, on the order of $10^{-3}$, agrees with the second-order of the schemes at the actual mesh size of $\frac{\aout-\ain}{N-1}=\frac{1}{30}$. The accuracy can be tuned by changing the quadrature and differentation schemes.

\section{Concluding remarks}

We have reformulated the Green kernel appearing in potential problems to circumvent the singularity and, at the same time, properly account for it. As a consequence, the gravitational potential of any celestial body, regardless its shape and matter density distribution, becomes directly accessible through two "classical" volume integrals, followed by a partial derivative. The method is applicable to three-dimensional, fully inhomogeneous systems, as well as to surface and line distributions. It is especially efficient inside distributions where most approaches exhibit a real practical complexity, converge very slowly, or produce spurious errors..
  The presence of regular kernels ensures that the method is stable and easy to implement. This should encourage modellers to abandon various integration techniques that do not ``faithfully'' reproduce the Newtonian character of the potential and forces. In the context of discs for instance, this method appears to be a real alternative to softened gravity, which remains a free parameter, non-Newtonian theory. As stressed, it is probably possible to determine other cool kernel/hyperkernel pairs (for instance, by considering the hyperkernel as a radial/angular gradient), but the one presented in the body of this paper seems the simplest one. It is in particular well-suited to axially symmetric configurations.

This study needs to be continued in several respects, including the analysis of the mathematical properties of the cool kernel and hyperkernel and their physical meanings, as well as the derivation of the equivalent kernel in other systems of coordinates (e.g. cartesian and spherical coordinate systems). In addition, it would be interesting to expand the two kernels in Eq.(\ref{eq:newpsi-Z}) in series, and compare their properties with the expansion of the Green function in Legendre polynomials, inside as well as outside the body. The cool kernel/hyperkernel pair is also interesting as a new starting point to generating various kinds of approximations. Apart from the astrophysical context where there are so many applications about gravitation, this technique is also transposable to other kinds of problems involving improper integrals. This can be, for instance in electromagnetism, the determination of the potential vector $\vec{A}$ and associated magnetic field induced by current densities \citep{jackson98,ct99}. These points will be touched on in forthcoming papers. 

\begin{acknowledgements}
It is a pleasure to thank J. Braine, M. Gazeau, F. Hersant, and A. Trova as well as the second referee for comments and suggestions about infinitely flat systems. J.-M. Hur\'e is grateful to the CNU, section 34, for supporting a six-months full-time research project through CRCT-2011 funding delivered by the MESR.
\end{acknowledgements}

\bibliographystyle{aa}

\end{document}